\documentclass[12pt]{article}

\usepackage{psfrag}

\usepackage{graphicx}
\begin{document}

\title {Coulomb breakup effects on the optical potentials of  weakly bound nuclei}

\author {Awad A. Ibraheem$^{(a,b)}$\footnote {awad@df.unipi.it}  
     and  Angela Bonaccorso$^{(a)}$\footnote {bonac@df.unipi.it} \\
{\small $^{(a)}$ Istituto Nazionale di Fisica Nucleare, Sezione di
Pisa, }\\ {\small and Dipartimento di Fisica, Universit\`a di Pisa,}\\ {\small Largo Pontecorvo 3, 56127
Pisa, Italy.}\\
{\small  $^{(b)}$ Physics Department, Al-Azhar University, Assuit 71524, Egypt.}}

   \maketitle
   
\begin{abstract}

The optical potential of halo and weakly bound nuclei has a long range
part due to the coupling to breakup  that damps the   elastic
scattering angular distributions. In order to describe correctly the breakup channel in the case of
scattering on a heavy target, core recoil effects have to be taken into account.  We show here that
core recoil and nuclear breakup of the valence nucleon can be consistently taken into account. A
microscopic absorptive potential is obtained within a semiclassical approach and its characteristics
can be understood in terms of the  properties of the halo wave function and of the reaction mechanism.
Results for the case of medium to high energy reactions are  presented.
\end{abstract}
 \newpage
\vspace{1em}
\section{Introduction}

Since the advent of Rare Isotope Beams (RIBs) \cite{ta}
elastic nucleus-nucleus scattering with a radioactive projectile \cite{exp} is
a reaction which has been studied to a large extent in the attempt to
find characteristics that would be typical for a weakly bound nucleus
and would help understanding new phenomena such as the existence of  the halos  \cite{hjj}. 
It has been established that for  halo projectiles breakup 
of the valence particles is responsible for a damping in the elastic angular distribution 
starting from around 
5$^o$ at medium to high energies.
 
All theoretical methods used to describe the above mentioned reactions,
require at some stage of the calculation the knowledge of the
nucleus-nucleus optical potential. The optical potential is the
basic ingredient for the description of elastic scattering, but it is
important also in transfer and breakup calculations, since one needs to take into
account the core quasi-elastic scattering by the target while the valence neutrons are transferred or
 breakup. For example in the case of some two-neutron halo nuclei such as $^{14}$Be or
$^{11}$Li, their cores of $^{12}$Be and
$^{9}$Li are themselves weakly bound nuclei. For multinucleon transfer reactions planned in order
to obtain heavy exotic nuclei it will be important to have the appropriate optical potentials which
include breakup and which should be used in the intermediate steps of the reaction.  In the charge
exchange  reaction $^{11}B(^{7}Li,^{7}Be)^{11}Be$  the halo nucleus-nucleus optical potential
necessary to describe the final channel  \cite{cap} has a volume part obtained
with a double folding plus a very diffuse surface term fitted
phenomenologically to reproduce the final channel angular distribution. The effect of the surface potential
 reduced the absolute cross sections by about 50\%\cite{cap}, in accordance with the experimental data and it can be
interpreted as due to  the halo breakup.

Microscopic optical potentials for a halo
projectile have already been studied by many authors,  and a review of the
present situation can be found in \cite{af,pv}. One of these methods consists in  starting  from a phenomenologically
determined core-target potential and then to add the effect of the breakup of the halo neutron. This
process leads to adding a surface part to the core-target potential.
This new  surface peaked optical potential has been seen to have a
quite long range which reflects the properties of the long tail
of the halo neutron wave function. Such kind of potentials are often called
dynamical polarization potentials \cite{pv}-\cite{so}.

In a recent  contribution we proposed a new approach \cite{af} to the calculation of
the imaginary part of the optical potential due to nuclear breakup. It is based
on  a  semiclassical  method described  by Broglia and
Winther  in \cite{bwb,bpw} and used also by Brink and collaborators \cite{bpb,sb} to calculate
the surface optical potential due to transfer and on the Bonaccorso and
Brink  model for transfer to the continuum reactions \cite{bb}-\cite{abb}, the idea being that breakup
is a reaction following the same dynamics as transfer but leading mainly to continuum final states when
the incident energies per nucleon are higher than the average nucleon binding energy.  The
calculations were almost completely analytical and a simple, approximated formula was obtained which  helped understanding the origin
of the long range nature of the potential and its dependence on the incident
energy as well as on the initial neutron binding energy. The characteristics of
our potential were consistent with those of   potentials obtained with other
methods, in particular  the eikonal method of
Canto et al.\cite{cd} and application to the description of experimental data
were encouraging \cite{af}.

However it is very well known that for heavy targets recoil effects which give rise to the so called
Coulomb breakup, are important and actually dominant for a neutron halo \cite{carlos,bbb,kai,nak} in the breakup cross section
and therefore one wonders what would be their effect on the elastic scattering. Some work has already
been published in order to calculate the optical potential due to Coulomb breakup at low 
\cite{may},\cite{mack} or
intermediate energies \cite{carlos1,carlos}. However paper \cite{carlos} deals with proton halo breakup
which, as it has been demostrated in Ref.\cite{bbb} has to be treated with great care when compared to
neutron breakup. Therefore the methods used in
\cite{carlos} to include breakup in the optical potential are not expected to be applicable to the
neutron breakup case. We will show in this paper that the method used in Ref.\cite{af} can be used in
the case of Coulomb breakup as well and that the corresponding formalism, which is appropriate to
reactions performed at medium to high energies, well above the Coulomb barrier, is consistent with and
joins continuously to the formalism used by other authors \cite{may} at lower energies.

\section {Theory}

The method we use here is the same as in \cite{af} and it is based on the extraction of an optical
potential from the calculation of a phase shift.

 The elastic scattering probability is
$P_{el}=|S_{NN}|^2$,  given in terms of the nucleus-nucleus S-matrix.
We know that
 \begin{equation}|S_{NN}(b)|^2=e^{-4 \delta_I(b)}.\label{a}\end{equation}
In a semiclassical approximation \cite{bwb}, the imaginary part of the nucleus-nucleus
phase shift
$\delta_I$ is related to the imaginary part of the optical potential by

\begin{equation}\delta_I(b)=-{1\over 2\hbar}
\int_{-\infty}^{+\infty}\left(W_V({\bf r}(t))+W_S({\bf r}(t))\right)dt
\label{b}\end{equation}
where the volume potential is responsible for the usual inelastic core-target
interaction, while the surface term takes care of the peripheral reactions like
transfer and breakup.
${\bf r}(t)={\bf b_c}+vt$ is the classical trajectory of relative motion for the
nucleus-nucleus collision.

According to \cite{af} the surface optical potential
$W_S({\bf r}(t))$ due to
 breakup can be related to the breakup probability by
\begin{equation}
\int_{-\infty}^{+\infty}W_S({\bf r}(t))dt=-{\hbar\over
2}P_{b_{up}}(b_c)\label {1} \end {equation}
where  $P_{b_{up}}
=\sum_ip_i
$ are the breakup probabilities in
the various channels $i$. 
    In order to obtain the surface imaginary potential Eq. (\ref{1}) should be calculated
as an identity in the distance of closest approach, which amounts to require that $ W_S(r)$
be a local, angular momentum independent function. We remind the reader that  since
we are using a semiclassical method, the non locality, which is in
principle a characteristic of microscopic optical potentials
 has been transformed into an energy dependence \cite{bbopt}.
    
    Eq.(\ref{1})  can also be derived in a straightforward way from the  time
dependent scattering Schr\"odinger equation for the elastic channel probability density function  in
presence of a complex potential
\cite{sb,sh}. In the traditional formulation the index (i) stands for stripping and pickup to bound states
and in Ref.\cite{af} we extended it to hold for breakup reactions in which the final neutron state is in
the continuum. Nuclear breakup of both absorptive and diffractive type was  included and here we will
include also Coulomb breakup. The justification of the use of Eq.(\ref{1}) to calculate the imaginary
potential due to nuclear breakup was simply given by the analogy between breakup and transfer as  expressed
by the transfer to the continuum model introduced in Refs.\cite{bb}-\cite{abb}. There it was shown that the
formalism for transfer to bound states goes over transfer to the continuum in a natural way if the
kinematics of the reaction is taken into account correctly within a time dependent approach which
ensures neutron energy conservation.

On the other hand Broglia and
Winter  in \cite{bwb,bpw} pointed out the fact that the same formalism could be extended to include
core recoil. In the case of breakup of weakly bound nuclei we have shown in Ref.\cite{jer,jer1} that core
recoil is responsible for the  Coulomb breakup and that nuclear and Coulomb breakup give rise
to negligible interference effects. Therefore we argue here that the imaginary part of the optical
potential due to Coulomb breakup can also be calculated by Eq.(\ref{1}) where now  one of the $p_i$
probabilities will be that of Coulomb breakup of the valence nucleon.

Then, using  Eq.(\ref{b}) and (\ref{1}), in (\ref{a}) the nucleus-nucleus S-matrix, in
the case of a halo projectile, can be written as

\begin{equation}|S_{NN}|^2=|S_{CT}|^2 e^{-P_{b_{up}}}\label{s}\end{equation}
where $S_{CT}$ takes into account all core-target interactions while the term
$e^{-P_{b_{up}}}$ depends only on the halo neutron breakup probability.
For a halo nucleus at high incident  energy the transfer
probability is going to be much smaller than the breakup probability, therefore the surface
potential has been identified here with the breakup potential.

Now we discuss the hypothesis leading to Eq.(\ref{s}). They have been already discussed in
Ref.\cite{af} but we report them here too for the sake of completness.

  In this paper  we are concerned  with  reactions  performed at  energies well above the
Coulomb barrier where many inelastic channels open at about the same distance of closest approach.
The effect of the breakup is most important at large distances  of closest approach ($b_c>R_s$),
 where it represents the dominant reaction mechanism. If the breakup probability is
needed at smaller impact parameters, then the values  calculated by
perturbation theory, have to be multiplied by the core survival probability, as discussed
in Eq.(V.8.1) of Broglia and Winther and also used in relation to halo breakup by several
authors.
 The effect of all inelastic channels $n$ different
from the one we are interested in, can be taken into account by
introducing a damping factor $P_0$. Therefore the breakup probability
$P_{b_{up}}$ at all distances can be defined as

\begin{equation} P_{b_{up}}=p_{b_{up}}\prod_n(1-p_n)\approx p_{b_{up}}
\exp(-\sum_n p_n)=p_{b_{up}}P_0\label{2}\end{equation}
Each elementary inelastic probability $p_n$ and breakup probability
$p_{b_{up}}$ is  small and $p_{b_{up}}$ in particular, can be calculated in
time dependent perturbation theory, as done in
\cite{bb}. In reactions with halo projectiles the damping factor $P_0$ has also
been referred to as the core survival probability
after the halo breakup or as the core elastic scattering probability. The
  breakup probability Eq.(\ref{2}) integrated over the core-target impact parameter $b_c$  has been
widely used in the literature to get 
 breakup cross sections.

In this paper we will treat both nuclear and Coulomb breakup as independent process with the formalism
used in Ref.\cite{jer}, namely we will calculate nuclear breakup in the eikonal approximation and
Coulomb breakup in first order perturbation theory. In Ref.\cite{jer,jer1} we showed that this is
appropriate for a one neutron halo since the interference effects are small and the higher order effects in Coulomb
breakup are negligible. This has been confirmed by recent experimental data \cite{nak}. The case of a proton halo or a
two-neutron halo  might need the full all-order approach.

Then the total breakup probability will be given by
\begin{equation}
{p_{b_{up}}(b_c)}=p^N_{b_{up}}(b_c)+p^C_{b_{up}}(b_c).\label{anc}
\end{equation}

\subsection {Nuclear breakup}
The optical potential due to nuclear breakup was extensively discussed in Ref.\cite{af}. We report
here on some of the most important results which will help us also constructing the potential for the
Coulomb breakup channel.

The nuclear breakup probability given by the eikonal model is

\begin{eqnarray}p^N_{b_{up}}(b_c)
&\approx& {\frac{C^2S}{2\pi}}\int
{\frac {d \varepsilon_f}{\hbar v}}\int  d{\bf b_n}(|1-S({\bf b_n})
|^2+1-|S({\bf b_n)}|^2)\nonumber \\& \times & {1\over (2l+1)}\sum_{m} |\bar \phi_{l,m}({\bf b_n-b_c},k_1)|^2
\label{dpde}\end{eqnarray}
where $C^2S$ is the spectroscopic factor for the initial state. 
$\bf {b_c}$ and ${\bf b_n}$ are  the core-target and the neutron-core impact parameters respectively.
The calculation of the nuclear breakup probability is done here in a reference frame with the center at the origin of the
target. It is important to remark that the above expression takes into account to all orders the neutron target
final state interaction via an eikonal S-matrix. In this way neutron elastic
scattering and absorption on the target are treated consistently. The integrand in Eq.(7) is surface peaked and goes rapidly to zero at the interior of the projectile potential 
(cf. Fig.5 of Ref.\cite{bobe}). This is  because of the natural cuts introduced by
 $S( b_n)$ and because of the large values of $b_c$. Thus the initial wave function is needed only at large radii and it can
be  approximated   by its asymptotic  form which is  an
Hankel function
   \begin{equation}\phi_{lm}({\bf r})=-i^lC_i\gamma h^{(1)}_l(i\gamma r)
   Y_{lm}(\theta,\phi), \,\,\,\gamma r >>1,
   \label{in}
   \end{equation}
where $C_i$ is the asymptotic normalization constant. Its one-dimensional Fourier transform reads
\begin{eqnarray}
{1\over (2l+1)}\sum_{m} |\bar \phi_{l,m}({\bf b_n-b_c},k_1)|^2
  &=&
{1\over (2l+1)} \sum_{m}|2 C_iY_{l,m}
(\hat k_1) K_{m}(\eta |{\bf b_n-b_c}|)|^2\nonumber \\  &\approx  &
C_i^2 {e^{-2\eta |{\bf b_n-b_c}|}\over 2\eta |{\bf b_n-b_c}|}P_{l}(X_i),
\label{2a}  \end{eqnarray} 
where $X_i=1+2k_1/\gamma$. The  divergency in the RHS of Eq.(\ref{2a}) is compensated in Eq.(\ref{dpde}) by the rapid decrease
of the terms depending on $S(b_n)$. 

Eq.(\ref{dpde}) is  the neutron breakup probability from a definite single
particle state of energy
 $\varepsilon_i$, momentum $\gamma=\sqrt
{-2m\varepsilon_i}/\hbar$, and angular momentum $l$  in the
projectile to all possible final continuum state of energy
$\varepsilon_f$ with respect to the target and  momentum $k_f=\sqrt {2m\varepsilon_f}/\hbar$. In our notation
$k_1=(\varepsilon_f-\varepsilon_i-{1\over 2}mv^2)/(\hbar v)$ and
$k_2=(\varepsilon_f-\varepsilon_i+{1\over 2}mv^2)/(\hbar v)$ are the
 $z$ components of the neutron momentum in the initial and final state,
respectively. $\eta^2=k_1^2+\gamma^2=k_2^2-k^2$ is the modulus square of the
transverse component of the neutron momentum. Recoil effects on the neutron are thus taken into account.
\begin{table}
\caption{Energy dependent optical model parameters for the  n-Pb
interaction, $a_{V} = 0.55fm$, $ a_{W} = 0.3fm$ , $ r_{V} = 1.25 fm$,  $ r_{W} = 1.26fm $.}
\begin{center}
\begin{tabular}{cccc} \hline
Energy&V&Wv&Ws\\
  MeV&MeV&MeV&MeV\\\hline\hline
20 &-44.8 &-2.03&-11.02\\
40 &-42.8&-4.02&-8.50\\
80 &-38.8&-6.08&-4.87\\\hline\hline
\end{tabular}
\end{center}
\label{potn}
\end{table}

 For the purpose of this paper the important thing is that in Eqs.(\ref{dpde}) the main dependence on the core-target
impact parameter $b_c$ is contained in the exponential factor
$e^{-2\eta |b_c-b_n|}$.  After the $b_n$ and  $\varepsilon_f$ integration, the breakup probability
 has still an exponential dependence on $b_c$ for impact parameters larger than the strong absorption radius. 
 This has been
shown already in Fig. 4 of Ref.\cite{bobe} and Fig. 1 of Ref.\cite{eb}. 
Also Eq. (\ref{dpde}) has a maximum 
in correspondence
to the minimum value of  $\eta=\gamma$. 
Therefore  the $b_c$-dependence of the breakup
probability
$p^N_{b_{up}}(b_c)$ will  be of the exponential form
$p^N_{b_{up}}(b_c)\approx  e^{-b_c/{\alpha}}$ with  $a\approx
(2\gamma)^{-1}$ where $\gamma$ is the decay length of the neutron
initial state wave function. We now assume at large distances, where $P_0=1$
the same exponential dependence for the absorptive potential due to nuclear breakup
$W^N_S(r)=W^N_0e^{-r/{a} }$. We assume also, as indicated earlier on, a straight line
parameterization for the trajectory ${\bf r}(t)={\bf b_c}+vt$, then
Eq.(\ref{1})  reads
\begin{equation}
\int_{-\infty}^{+\infty}W^N_S( b_c,z)dz=-{\hbar v\over
2}p^N_{b_{up}}(b_c).\label{1bis}
\end{equation} The LHS can be approximately evaluated as

\begin{eqnarray}\int_{-\infty}^{+\infty}
W^N_S(b_c,z)dz={W^N_0}\int_{-\infty}^{+\infty}e^{-(b_c+{z^2\over
2b_c})/a}dz  ={W^N_0}\sqrt{2\pi
b_ca}e^{-b_c/a},\label{09}\end{eqnarray}
where we assumed $b_c>>z$ in the second step.
Equating the RHS of Eqs.(\ref{1bis}) and (\ref{09})  and renaming the
distance $b_c$ as $r$ gives

\begin{equation}W^N_S(r)=
-{\hbar v\over 2}p^N_{b_{up}}(r){1\over \sqrt{2
\pi a r}}\label{9}\end{equation}

Eq.(\ref{9})  shows explicitly, as already discussed in Ref.\cite{af}, that the long range nature of the nuclear breakup
potential originates from the large decay length of the initial state wave
function. For a typical halo separation energy of  0.5MeV,
$a=(2\gamma)^{-1}=3.2fm$, while for a `normal' binding energy of 10MeV, $a=0.7fm$ as
expected. Therefore the parameter
$a$ will depend mainly on the projectile characteristics and not on the target.

\subsection{Coulomb breakup}

Coulomb breakup can  be taken into account as well,
following the formalism of
\cite{jer} and the first order perturbation theory probability reads
\begin{equation}p^C_{b_{up}}(b_c)
\approx {\frac{4C^2SC_0^2}{8\pi^{3}b_c^2}}\int d\varepsilon_k
{\frac{m_nk}{\hbar^{2}}}\int d\Omega_{k}
\left |\left(  \varpi K_{1}\left(  \varpi\right)  \frac{d}{dk_{x}
}+i\varpi K_{0}\left(  \varpi\right)  \frac{d}{dk_{z}}\right)  \tilde{\phi
}_{lm}\left(  \mathbf{k}\right) \right |^2.
\label{dpdec}\end{equation}
The constant $C_0={\beta}_{1}Z_{P}Z_{T}e^{2}/\hbar v$ is a dimensionless
interaction strength and $\varpi=(\varepsilon_k-\varepsilon_i) b_c/\hbar v=\omega b_c/v$ is the adiabaticity
parameter. There is part of the core recoil which goes to center-of-mass
motion.
This effect is  included in the effective charge parameter $\beta_{1}=m_n/m_P$ which is a relevant correction for
 light nuclei. The Coulomb breakup probability is best calculated in the projectile reference frame. Thus
$\varepsilon_k$ is here the neutron final energy with respect to the core. Obviously after integration over the final
energy both nuclear and Coulomb probabilities are independent on the reference frame they were calculated in. The
functions $K_{0}$ and $K_{1}$ are modified Bessel functions.

If the initial state wave function is approximated   again by its asymptotic  form Eq.(\ref{in} ),  then the general
form of the initial state momentum distribution is given by the three-dimensional Fourier  transform of
Eq.(\ref{in})
 \begin{equation} \widetilde {\phi}_{lm}({\bf k}) = 4\pi  C_i {{k}^l\over
\gamma^l(k^2+\gamma^2)}Y_{l,m} (\hat{k})
 \label{22}  \end{equation} where ${\bf {k}}\equiv (k_x,k_y,k_z)$ is a real vector.

In the
amplitude for Coulomb breakup  the Fourier
transform of the bound state wave function is well approximated by the Fourier transform Eq.(\ref{22})
of the corresponding Hankel function provided that $\gamma R < 1$ and
$kR < 1$ where  R  is the  radius of the neutron-core potential.  These conditions are
often satisfied in the Coulomb breakup of a halo nucleus.  For example
 for $^{11}$Be, $R\simeq 2.5 fm$, $\gamma=0.15fm^{-1}$ and the largest final momentum entering the calculation is
$|k|\simeq 0.3fm^{-1}$. For this range of final momenta the Fourier transform Eq. (\ref{22}) is larger by about 10\% than the Fourier
transform of the Woods-Saxon numerical wave function but it has a very similar shape. 
The assumption  of the asymptotic form is
expected to overestimate the
Coulomb excitation by about 20\%. The nuclear
excitation is better approximated since the Fourier transform is taken only in one dimention while $|\bf {b_n-b_c}|$ is always
well outside the neutron-core potential. There are however other incertitudes which affect the total probability values. 
One is 
the not perfectly known spectroscopic factor. The other is the value of the asymptotic normalization constant which depends on
the geometry used for the neutron-core potential. This last incertitude would be present even if the proper Woods-Saxon wave
functions were used. Also it is important to note that the 2s wave function has a node which gives a divergency in its Fourier
transform around
$k=0.55 fm^{-1}$. However for the situations studied in this paper the integrand in Eq.(\ref{dpdec}) has negligible values for
such large final momenta. Also using wave functions calculated in a square-well potential it is easy to show that for
$|\varepsilon_i|\leq 1$MeV and up to $l=2$ the contribution of the
internal part of the wave function to the full Fourier transform can be
neglected. Here the final state has been taken as a plane wave. We have shown in Fig. (8) of Ref.\cite{jer1} that using
a proper continuum wave function gives negligible differences when the initial state has $l=0$.

For a $l=0$ initial state which is appropriate for the halo breakup of $^{11}$Be for example,
and after integration of the angular variables, Eq.(\ref{dpdec}) reads

\begin{equation}{p^C_{b_{up}}}(b_c)
\approx {\frac{32}{3\pi}}C^2S\left(\frac{{C_0C_i}}{b_c}\right)^2\int d\varepsilon_k
{\frac{m_n}{\hbar^{2}k}}
\left (  \varpi^2 K_{1}^2\left(  \varpi\right)
+\varpi^2 K_{0}^2\left(  \varpi\right) \right){ k^4\over (k^2+\gamma^2) ^4}.
\label{dpdec2}\end{equation}

We show in Appendix A that inserting this result in Eq.(\ref{1}) leads to an expression consistent
with Eq. (12) of Andr\'es et al.
\cite{may}. Those authors have developed a semiclassical method to obtain a polarization potential for the Coulomb breakup channel at energies close to the barrier.   Our formalism can then be viewed as a natural extension of the formalism
of \cite{may} in the high energy, straight  line trajectory limit in which the eikonal model of the
phase shift Eq.(\ref{b}) is valid. Also our approach can be considered consistent to that of
Ref.\cite{canto} since those authors showed the consistency of their method
 to that of Ref.\cite{may}

\subsection{Extraction of the imaginary potential}

   \begin{figure}[h]
\begin{center}
\includegraphics[scale=.35,angle=0]{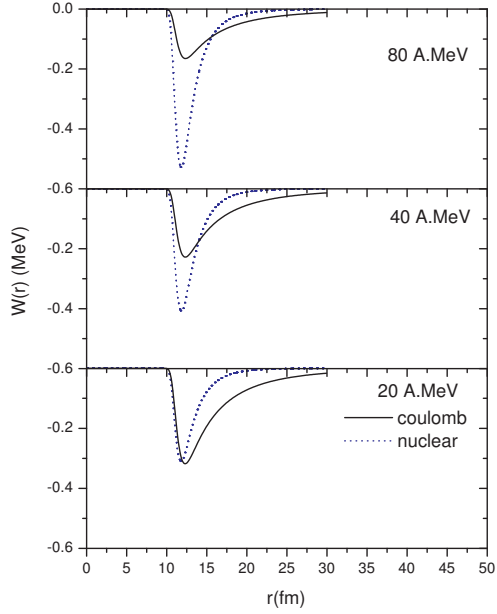}
\end{center}
 \caption{\footnotesize { Nuclear and Coulomb breakup potentials}}
\label{pot4l}
 \end{figure}

Eqs.(\ref{b}), (\ref{1}) and (\ref{s}) show that the
imaginary part of the phase shift and the modulus square of the S-matrix which takes into account the breakup channel can be simply obtained from the  breakup probability and it  would not be  necessary to know the actual form of the optical potential.
However to understand the physical origin of its characteristics and in order to use elastic scattering data for
structure studies it is useful to deduce an analytical form of it. To this goal  we have followed
the method outlined in Sec. 2.1. We started by calculating the total Coulomb  and nuclear breakup probabilities from the  numerical integration over the neutron final continuum energy of
Eqs. (\ref{dpde}) and (\ref{dpdec}). The neutron-target optical potential used here to calculate the neutron breakup
probabilities is from Ref.\cite{mahux} and it has the usual Woods-Saxon form for the real and imaginary volume
 parts and a Woods-Saxon derivative form for the surface imaginary part. The parameter values are given in Table 1.
 We then fitted the total probabilities to  a sum of exponentials. This allows a simple extraction of the potential
from the relation Eq.(\ref{09}) and also a transparent physical interpretation of the large diffusness.
\begin{equation}
p^N(b_c)= \sum_n A_n ~\exp {(-b_c/\alpha_n)},
\label{AQ1}\end{equation}
\begin{equation}
p^C(b_c)= \sum_n B_n ~\exp {(-b_c/\beta_n)},
\label{AQ2}\end{equation}
\begin{table}
\caption{Best fit obtained for the nuclear and Coulomb halo breakup 
probabilities Eqs.(\ref{dpde}),(\ref{dpdec}) for $^{11}$Be incident on $^{208}$Pb. The parameters
$\alpha_n$ and $\beta_n$ are in fm, while $A_n$ and $B_n$ are
dimentionless. Incident energies in A.MeV }
\begin{center}
\begin{tabular}{|c|ccc||ccc|c|}
\hline \hline &&{\bf Nuclear }&&&{\bf Coulomb }&&\\\hline
$E_{inc}$&20&40&80&20&40&80&\\\hline\hline
$A_1$&1860.38&303.23&339.45&9.672&3.922&1.734&$B_1$\\
$A_2$&19.73&7.91&1.91&1.558&0.6013&0.2549&$B_2$\\
-&-&-&-&0.1531&0.0587&0.0248&$B_3$\\
$\alpha_1$&1.1924&1.5334&1.5356&2.7273&2.9438&3.1046&$\beta_1$\\
$ \alpha_2$&2.5403&2.7918&3.4955&6.5189&7.4850&8.2713&$\beta_2$\\
-&-&-&-&15.6863& 19.9322& 24.1546&$\beta_3$\\\hline\hline
\end{tabular}
\end{center}
\label{tab1}
\end{table} 

The corresponding parameters are given in Table 2.
The parameters $\alpha_n$ and $ \beta_n$ obtained  show clearly that the exponential tail of the breakup probabilities have a very long range, in particular in the Coulomb breakup case.
To obtain  forms of the  potentials valid at all distances, we used then
 Eqs.(\ref{2}), (\ref{1bis}) and (\ref{09}), renaming again $b_c$ with $r$ such that
 \begin{equation}W^N_S(r)=
-{\hbar v\over 2}P_0\sum_n A_n ~\exp {(-r/\alpha_n)}{1\over \sqrt{2
\pi \alpha_n r}}\label{9a}.\end{equation}
\begin{equation}W^C_S(r)=
-{\hbar v\over 2}P_0\sum_n B_n ~\exp {(-r/\beta_n)}{1\over \sqrt{2
\pi \beta_n r}}\label{9b }.\end{equation}

 The core survival probability has
been parameterized as
\begin{equation}
P_0(b_c)=|S_{CT}|^2=\exp(-\ln2e^{[(R_s-b_c)/a_0]}),\label{pel}\end{equation}
where $a_0=0.6fm$ and $R_s=1.4 (A_P^{1/3}+A_T^{1/3})fm$ is  the strong absorption radius \cite{dmb}.

\section {Results}
    \begin{table}
\caption{Optical potential parameters for the bare  $^{10}$Be,  $^{11}$Be-$^{208}$Pb interaction. 
Radii are calculated with the  $R_i=r _{i} (
A_{P}^{1/3}+A_{T}^{1/3})$fm convention. }
\begin{center}
\begin{tabular}{|c|ccccccc|} \hline
Energy(A.MeV)&V&$r_{V}$&$a_{V}$&Wv&$r_{W}$&$a_{W}$&\\\hline
20 & -80&1.020&0.78&-66.7&1.110&0.39&\cite{buenerd}\\
40 &-70&0.920&1.04&-58.9&0.890&0.89&\cite{pot3} \\
80 & -36&1.087&0.78&-40.0&1.040&0.41&\cite{pot2}\\\hline \hline
\end{tabular}
\end{center}
\label{pot}
\end{table}

 In order to sample the quantitative accuracy of the simple analytical model presented
above we discuss now some numerical examples. The potentials we will discuss derive from
the breakup of the $2s_{1/2}$  state  of   $^{11}$Be, with separation energies
$0.5MeV$, asymptotic normalization constant $C_i=0.91 fm^{-1/2}$ and spectroscopic factor $C^2S=0.77$.
$C_i$ was obtained from a wave function calculated in a Woods-Saxon potential with radius parameter
 $r=1.25fm$, diffuseness $a=0.8fm$ and depth $V_0=-52.4MeV$, adjusted to fit the neutron separation energy.

\begin{figure}
\psfrag{sigmar\r}{$\sigma/\sigma _{R}$ }
 \psfrag{thetar\r}{$\theta$ $_{C.M}$ (deg)}
\psfrag{this bare potential fixed for 40 mev/n for all.\r}{.}
\includegraphics[scale=.35,angle=0]{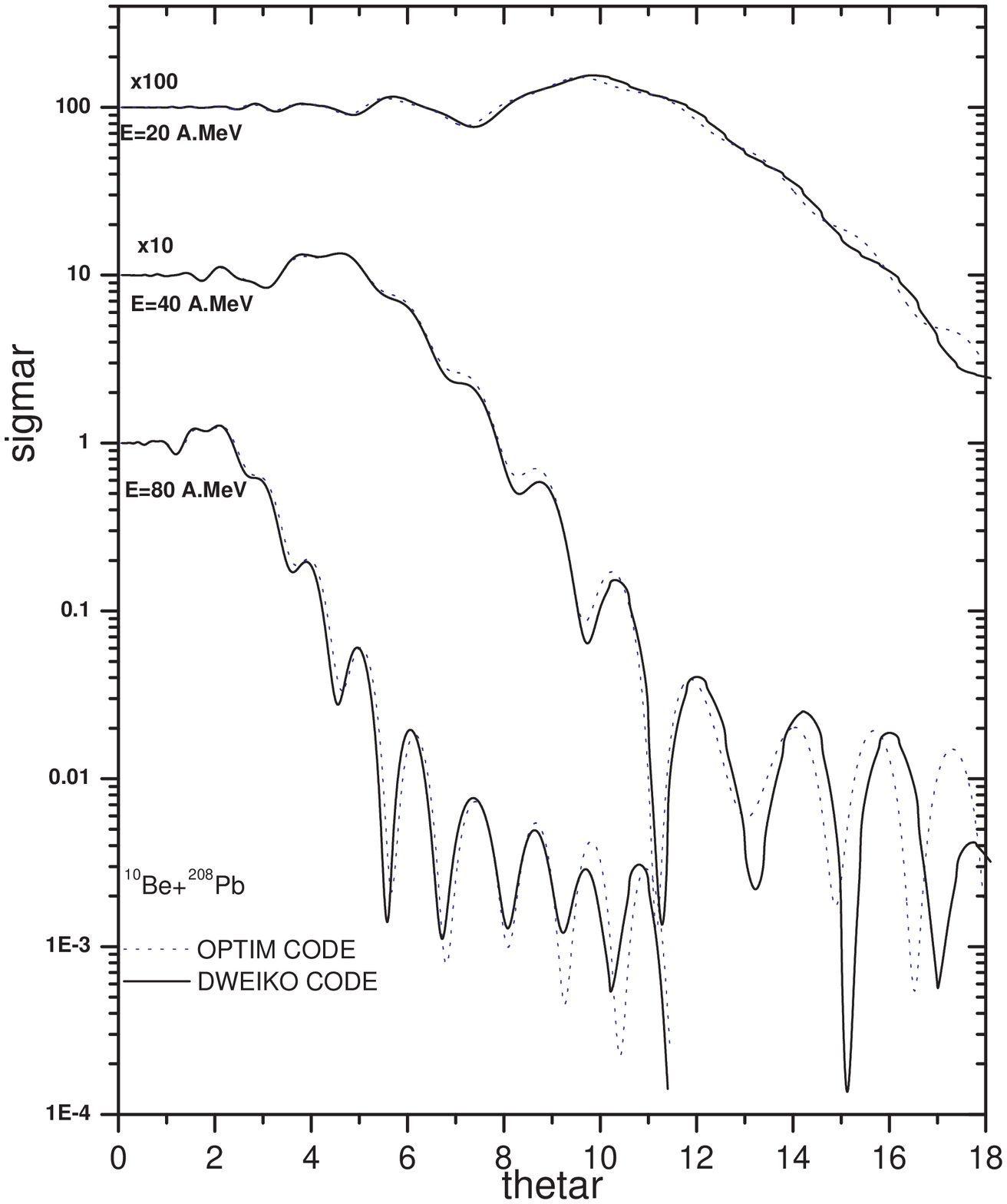}
 \centering \psfrag{sigmar\r}{ .}
\includegraphics[scale=.35,angle=0]{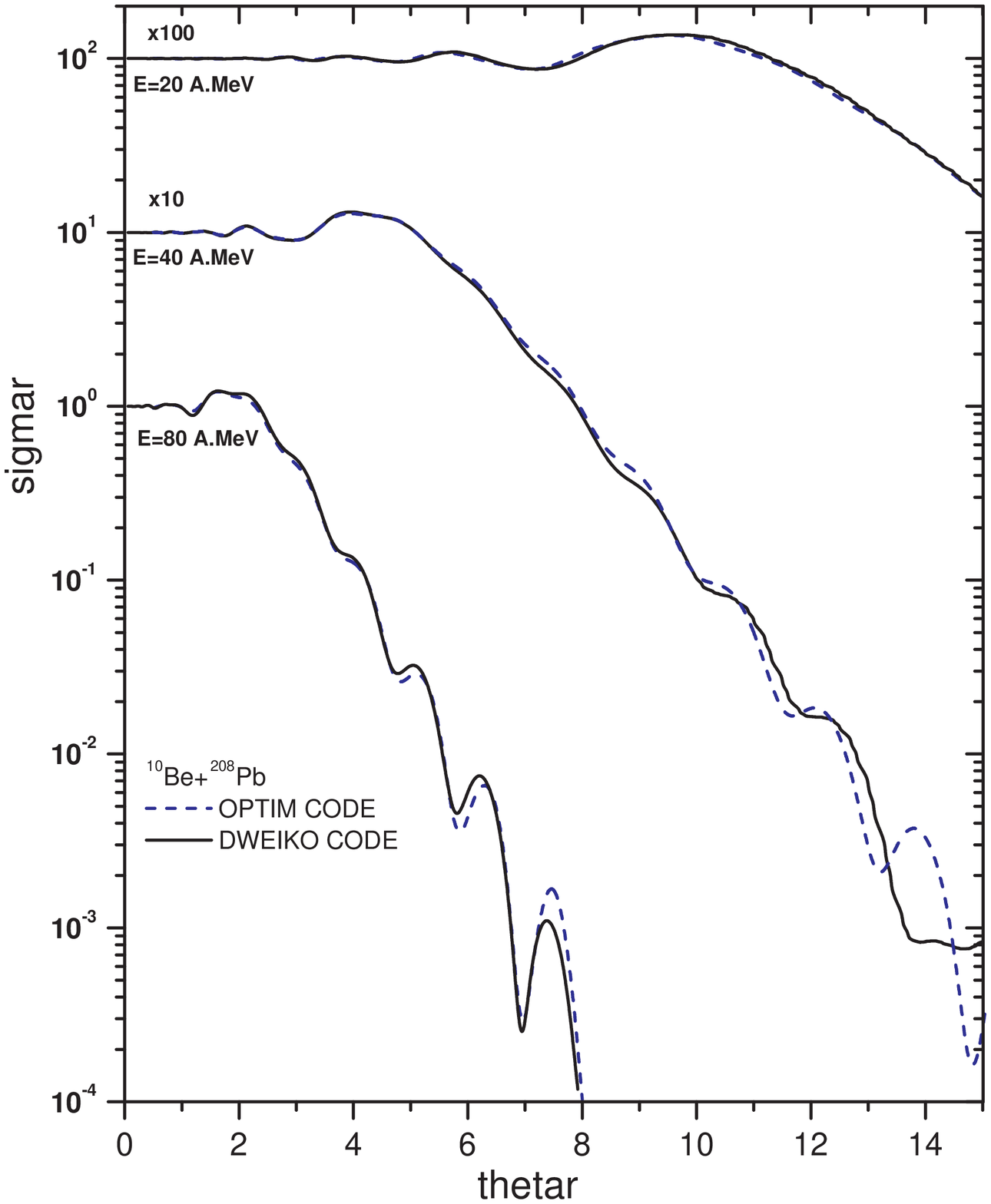}
\caption{\footnotesize {Comparison between the results of two  calculations according to the eikonal model (solid line) and the optical model (dotted line) at  20, 40 and 80 A.MeV for $^{10}$Be +$^{208}$Pb with the potential 
\cite{pot3}, right hand figure, and with the potential \cite{pot2}, left hand figure. }} \label{fcomparison}
\end{figure}

As we discussed already in detail in Ref.\cite{af} the diffuseness of the nuclear breakup potential is a
very important parameter. It depends weakly  on the incident energy and its large value between 2 and 3 fm is
due to the large decay constant of the neutron halo wave function and thus to the low separation energy.
In the case of the Coulomb potential  we see that the diffuseness $\beta_n$ is even larger, and furthermore increasing
with energy. This is clearly a reflection of the long range, energy dependent effects of  the Coulomb potential itself
and of the core-recoil it causes.

 From the diffuseness parameters values shown
in Table 2 we notice that at each incident energy the largest diffuseness values are
\begin{equation}\beta_3\approx ( \varepsilon_k^{max}-\varepsilon_i) /\hbar v\end{equation}
where $ \varepsilon_k^{max}=2\div 3MeV$ is of the order of  the largest continuum energy for which there 
is an appreciable breakup probability. This effect is clearly related to the behavior  
 at very large distances of the Bessel functions $K_{0,1} \approx e^{-\varpi}$ in Eq.(\ref{dpdec}).
 Therefore in the case of the  Coulomb breakup potential the
diffuseness value is related to the adiabaticity parameter and 
it depends  on the initial separation energy but also,  strongly, on the incident energy. 
Thus this potential is much
more  dependent from the reaction dynamics than the nuclear breakup potential.

\begin{figure}[ht]
\begin{center}
\psfrag{sigmar\r}{$\sigma/\sigma _{R}$ }
 \psfrag{theta\r}{$\theta$ $_{C.M}$ (deg)}
\includegraphics[scale=.45,angle=0]{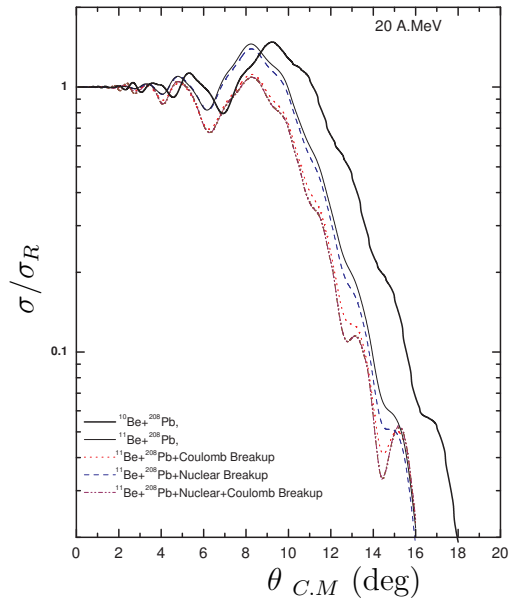}
\end{center}
 \caption{\footnotesize { Angular distribution for the elastic
scattering at 20 A.MeV of $^{10}$Be and  $^{11}$Be from $^{208}$Pb.  }}
\label{20total}
 \end{figure}

We show now in Fig.1 the
radial shapes of
the potentials calculated for the breakup from the
$2s$  state of $^{11}Be$ in the interaction with
$^{208}Pb$.   Results are given for three
laboratory incident energies:  E$_{inc}$=20 A.MeV,  40 A.MeV and
80 A.MeV. The solid lines are the Coulomb breakup potentials  while the dotted lines are the nuclear
breakup potentials.
  As discussed above  our results show that the potentials are energy dependent and the Coulomb breakup potential has a much longer range than the nuclear breakup potential.
   \begin{figure}[th]
\begin{center}
\psfrag{sigmar\r}{$\sigma/\sigma _{R}$ }
 \psfrag{theta\r}{$\theta$ $_{C.M}$ (deg)}
\includegraphics[scale=.35,angle=0]{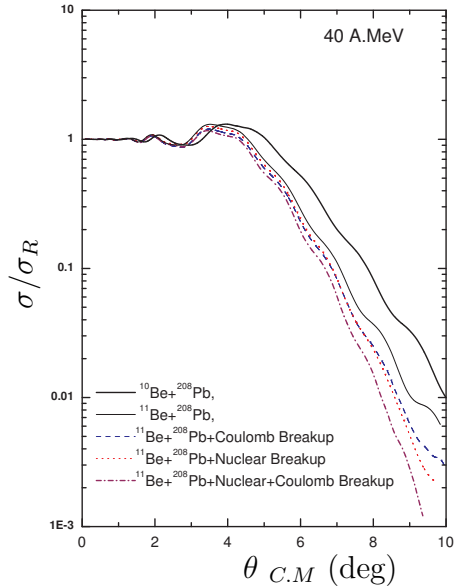}
\end{center}
 \caption{\footnotesize { The same as Fig. 3 but at  40 A.MeV }} \label{F10be40}
\end{figure}

The optical potential has one of its most interesting application in
the calculation of elastic scattering angular distributions. Since
according to Eqs.(\ref{b})  and (\ref{1bis}) our phase shifts for
the breakup channels are calculated in the eikonal model, it seems
appropriate to calculate the angular distributions within the same
model. Therefore we first  show in Fig. 2 a comparison of
calculations made with the optical model code OPTIM \cite{nobby} and
with our modified  version of the eikonal code DWEIKO \cite{dweiko}
for the reaction  at the same energies at which
potentials have been given in Fig.1. For the lowest energy DWEIKO contains a further recoil 
correction obtained by  
shifting the integration impact parameter, as
explained in Ref.\cite{dweiko}.
 The  optical
model parameters for the volume parts of the bare potential are given in Table 3. Since there are no data in the literature for the $^{10}$Be+$^{208}$Pb system at the incident energies we are concerned with we have tried to use two modified potentials obtained from fits of  $^{12}$C+$^{208}$Pb \cite{pot3} and $^{4}$He+$^{208}$Pb \cite{pot2} data.
  It is clear that for both potentials the eikonal model works very well up to 10$^o$.  At very high incident energies
when an optical model code could have difficulties in handling the large number of partial waves necessary
for heavy ions, an eikonal calculation is much simpler and equally accurate. A serius problem for the check of accuracy of theoretical models and related numerical applications of the
kind of problems discussed in this paper is the lack of experimental data and of reliable bare potentials for
the core-target interaction. Therefore the calculations discussed in the following have to be considered as
exploratory.

  In Figs. 3, 4 and 5 we show then the eikonal model angular distributions for $^{10}$Be+$^{208}$Pb
by the thick solid   line. The thin solid line is for $^{11}$Be+$^{208}$Pb with the bare volume potential of Table 3.
 The dashed and dotted lines  include the Coulomb  breakup and the nuclear breakup potentials respectively, while the
dot-dashed line includes both. As expected in most of the angular range the Coulomb breakup  reduces the elastic scattering
more than the nuclear breakup. 
 
\begin{figure}[th]
\begin{center}
\psfrag{sigmar\r}{$\sigma/\sigma _{R}$ }
 \psfrag{theta\r}{$\theta$ $_{C.M}$ (deg)}
\includegraphics[scale=.35,angle=0]{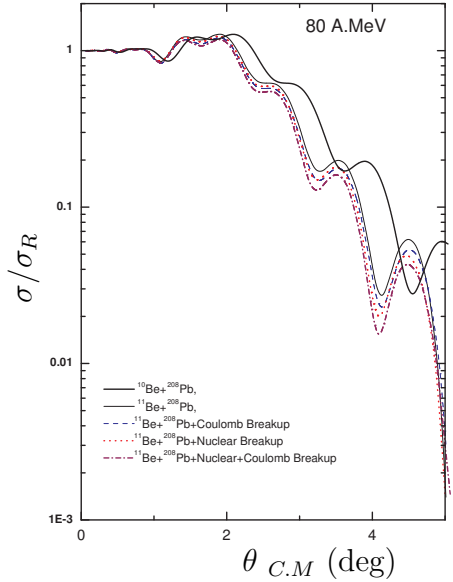}
\end{center}
\caption{\footnotesize {The same as Figs. 3 and 4 but at  80 A.MeV }}
 \label{F10be80}
\end{figure}

\begin{figure}[th]
\begin{center}
  \includegraphics[scale=.35,angle=0]{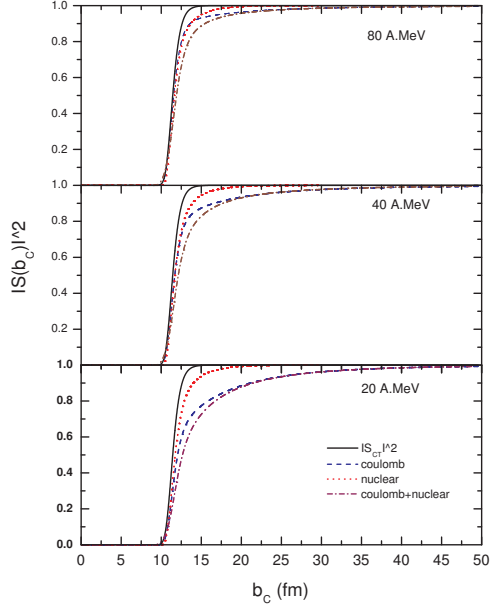}
\end{center}
\caption{\footnotesize {Scattering matrix with and without breakup potentials contributions }}
 \label{sct}
\end{figure}

Finally we discuss another important effect namely how the elastic scattering total probability
changes as a function of the impact parameter or angular momentum when there is a large breakup. 
 In Fig.6 we show with the solid line the core-target S-matrix
$S_{CT}$. Because of the incertitudes in the bare optical potentials
used for the angular distribution calculations we prefer here to use
for 
 $S_{CT}$  a
unique  form given by Eq.(\ref{pel}).
 The nucleus-nucleus
S-matrix
$S_{NN}$, shown by the dot-dashed line, is calculated then from Eq.(\ref{s}),  which contains
the effect of the halo breakup.  
 At a
fixed impact parameter  the effect of the breakup is to reduce the elastic probability given by the modulus
square of the S-matrix.  The unitary limit is attained at much larger
$b_c$-values and the reaction cross section receives significant contributions from a large range of impact
parameters. The reduction is obviously more pronounced at the impact parameters larger than the strong
absorption radius.
  The value of the strong absorption radius, defined as $|S_{NN}(R_s)|^2={1\over 2}$  changes appreciably, increasing of about 1 fm at 20 A.MeV, 0.8 fm at 40 A.MeV
  and 0.6 fm at 80 A.MeV. The large increase is mainly due to the Coulomb breakup effect. This can be seen by looking at the dotted and dashed lines which represent $S_{NN}$  when only the nuclear or the Coulomb breakup effect is included, respectively.

Another significant
effect of the imaginary surface potential
is seen in the  reaction cross sections obtained from the $S_{NN}$ and given in  Table 4. We show also a number of other cross section  values necessary to justify the phenomenological inputs of our calculations and their consistency.  The first two rows give the total cross 
sections  for the systems  $^{10}$Be + $^{208}$Pb and $^{11}$Be + $^{208}$Pb respectively, calculated with  the bare optical
potentials of Table 3. The cross sections in  the next three rows contain also the effect of the nuclear breakup, of the
Coulomb breakup  and of both at the same time. All these cross sections have been calculated in the eikonal approximation  with
the code  DWEIKO. There is 
 an increase of the order of 500$\div$600 mb with respect to the bare (no breakup) optical
potential when the nuclear breakup potential is taken into account. When the Coulomb breakup is included we find an increase of about 1.6 b to 4.4 b depending on the incident energy. The total increase in the reaction
cross section is very large,  corresponding to a variation of a factor from 2 to 3 at some energies. Similar increases in the cross sections have been found in Ref.\cite{mack}. These values are consistent with  the nuclear and Coulomb breakup cross sections obtained from the integration over the core-target impact parameter of the probabilities  Eqs.(\ref{dpde}) and (\ref{dpdec}) including the core survival probability Eq.(\ref{pel}) and given in the next two rows.  Finally  the free particle n-Pb elastic, inelastic and total cross sections obtained in the eikonal model by using the potentials of Table 1 are given. The last row contains the experimental free particle cross sections.  As expected at 20MeV our free particle cross sections are still not very close to the experimental values, but already at 40MeV the eikonal approximation works very well.
   
  As discussed already in \cite{af},  the relation between the total reaction cross section and the breakup cross section, can be understood by expanding the exponential in Eq.(\ref{s}) to first order in
$P_{b_{up}}$ and integrating over the impact parameter $b_c$. One immediately finds

\begin{eqnarray}
1-|S_{NN}(b_c)|^2&\approx&
1-|S_{CT}(b_c)|^2e^{-P_{b_{up}}(b_c)}\nonumber \\
&= &
1-|S_{CT}(b_c)|^2+\nonumber \\ &&+|S_{CT}(b_c)|^2( p^N_{b_{up}}(b_c)+p^C_{b_{up}}(b_c))
\end{eqnarray}

\begin{eqnarray}\sigma_{NN}& = &2\pi\int b_c db_c
\left (1-|S_{NN}(b_c)|^2\right ) \label{crt1} \\&\approx &
\sigma_{CT}+\sigma^N_{b_{up}}+\sigma^C_{b_{up}}.\label{crt2} \end{eqnarray}

The fact that the cross sections values of Table 4 are in very good agreement with the relations contained in
Eqs.(\ref{crt1}) and (\ref{crt2}) is a proof of the accuracy of the hypothesis Eq.(\ref{s}) for the nucleus-nucleus
S-matrix and of the separation of the imaginary potential into a volume and a surface term, with the surface term
identified with the halo breakup potential.

\begin{table}
   \begin{center}
\caption{Cross section values (in mb) discussed in this paper. 
  See text  for details. }
\begin{tabular}{|c|c|ccc|} \hline \hline
&Energy(A.MeV)&20&40&80\\\hline
%$\sigma_{NN}(^{10}Be)
{\footnotesize$^{10} Be+^{208}Pb$ total cross section}&$\sigma_{CN}$&2362&2897&2537 \\
{\footnotesize$^{11} Be+^{208}Pb$ total cross section with bare potential}&$\sigma_{NN}$&2644&2969&2593 \\
{ \footnotesize $\sigma_{NN}$  including nuclear breakup}&$\sigma^{N}_{NN}$&3185&3471&3060 \\
{\footnotesize $\sigma_{NN}$ including coulomb  breakup}&$\sigma^{C}_{NN}$&6685&5477&4152 \\
{\footnotesize $\sigma_{NN}$ including coulomb  and nuclear breakup}&$\sigma^{N+C}_{NN}$&7074&5900&4584 \\
{\footnotesize nuclear halo breakup }&$\sigma^{N}_{b_{up}}$&615&551&490\\
{\footnotesize Coulomb halo breakup}&$\sigma^{C}_{b_{up}}$&4280&2638&1570 \\
{\footnotesize n-$^{208}Pb$ elastic free particle } &$\sigma^{el}$&2554&2623&2467\\
 {\footnotesize n-$^{208}Pb$ inelastic  free particle }&$\sigma^{in}$&2018&1902&2097\\
{\footnotesize n-$^{208}Pb$ total free particle } &$\sigma^{Tot}$&4548&4437&4867\\
{\footnotesize n-$^{208}Pb$ experimental} &$\sigma_{exp}$ \cite{prc47}&5888&4392&4817\\\hline\hline
\end{tabular}
\end{center}
\label{XXXX}
\end{table}

\section{Conclusions}
 In conclusion we have presented a simple analytical method to obtain
the surface component of the  imaginary part of the nucleus-nucleus optical potential  to be used in the elastic scattering calculations between a halo or weakly bound nucleus and  a heavy target. The surface potential is due to Coulomb and
nuclear breakup. The
main purpose here was to relate the characteristics of the potential to the special properties of the
breakup channels for weakly bound nuclei. At high incident energy ($>$ 20A.MeV) the evaluation of the potentials amounts
in fact just to the calculation of the breakup probability as already shown in Ref.\cite{af}. If breakup
from core excited states is to be included, then it suffices to sum up the relative probabilities
according to Eq.(\ref{1}).

The method to include Coulomb breakup is an extension of that previously used to calculate microscopically
the effect of transfer and nuclear breakup channels on the imaginary potential. The shape of the
surface imaginary  potential and its
parameters are determined univocally by the shape of the breakup probability.  An
interesting result is that the diffuseness of the potential for the nuclear breakup  reflects the decay length of the
valence neutron wave function  and therefore it depends mainly upon the projectile
structure, but not so much on the reaction dynamics. For the Coulomb
breakup there is instead a strong dependence on the dynamics which leads to an increase of the diffuseness value when
increasing the incident energy. The strength  is also  energy dependent.  Analytical calculations contained in
appendix A  have shown that the potential proposed here is consistent   with and can be viewed  as an extension to
high energy, of  other theoretical models developed at energies close to the Coulomb barrier \cite{may}. The numerical
calculations on the other hand show consistency with the works of other authors
\cite{mack} for similar, light halo systems.   Furthermore we have
given an explicit justification for the long range of the polarization
potential.

   From Feshbach  \cite{fesh} formalism it is known that related to  the imaginary potential, which  comes from the second order, 
complex term, there is  also a real correction to be added to the first order term, the  folding
potential.  However it has been shown by the authors of Ref.\cite{may} that the Coulomb excitation polarization
potential is purely imaginary at high energy. On the other hand for the nuclear breakup potential, we have shown in
Ref.\cite{af} that the second order real correction is small and negligible, while other calculations \cite{yab,jat2}
for
$^{11}Li$ projectile found a not so small real polarization potential. To make sure that the second order real
correction is  negligible also for heavy targets such as $^{208}$Pb, we have added phenomenologically a real repulsive part to our microscopically calculated imaginary
potential,  with the same exponential form, same diffuseness and variable strength, but we found no noticeable
differences in the angular distributions. Two open questions remain then to be addressed  for future work. One is which
   observable will be more sensitive to the real correction term. The other is a systematic  microscopic calculation of such a term for a number of different systems and energies.

   {\bf Acknowledgments}

We are very grateful to Carlos Bertulani for providing and helping us using  his code DWEIKO
 and for his very useful remarks on our manuscript. We wish to thank also David Brink and Alvaro Garcia-Camacho for
discussions and Stefan Typel for providing us the good parameters of his potential. One of us (A.A.I) is very grateful to the Italian Ministry of Foreign Affairs
for a one year grant under the scheme ' Inter-University Cooperation. Scholarships and Youth Exchange Programmes'.

   \appendix\section{Connection to low energy approaches}

In our formalism the imaginary part of the optical potential due to Coulomb breakup is given by
\begin{equation}
\int_{-\infty}^{+\infty}W_S({\bf r}(t))dt=-{\hbar\over
2}p^C_{b_{up}},\label {a1}
\end{equation}
while  Eq. (12) of Andr\'es et al. \cite{may} reads
\begin{equation}
\int_{-\infty}^{+\infty}W_S({\bf r}(t))dt=\hbar\alpha.\label {a2}
\end{equation}
Where both $p^C_{b_{up}}$ and $\alpha$ depend on the incident energy and on the classical trajectory
of relative motion. In our case such a trajectory is a straight line since our approach applies to
incident energies well above the Coulomb barrier.

The two approaches are consistent if $\alpha \to
-{p^C_{b_{up}}/ 2}$ in the high energy limit.
To show that this is true we write explicitly the expression in Ref. \cite{may}
\begin{equation}
\alpha=-{\pi \over 9}\left ( {Z_{T}e\over \hbar v a_c}\right)^2\int d\varepsilon_k\left(I^2_{11}(\varpi)+I^2_{1-1}(\varpi)\right){dB(E1,\varepsilon_k)\over
d\varepsilon_k}\label{alpha}\end{equation} where
$I_{1\pm1}$ are the well known Coulomb integrals \cite{wa} which can be expressed in terms of Bessel
functions of imaginary order.  However according to Eq.(28) of
Ref.\cite{b1} in the high energy limit
\begin{equation}I(E1,\pm 1)=I_{1\pm1}={2a_c\over b_c}\varpi K_1(\varpi),\label{I}\end{equation}
where $a_c$ is the Coulomb length parameter and now $K_1$ is an ordinary Bessel function of real
index.

On the other hand considering Eq.(\ref{dpdec2}) for $p^C_{b_{up}}$ we remark that it is well known and
shown for example in Fig.1 of Ref.\cite{he} that
$\varpi^2 K_{0}^2\left( \varpi \right) $ is much smaller than $\varpi^2 K_{1}^2\left(  \varpi \right ) $ for values
of $\varpi \approx 0.1$, which happens for heavy ions at high energies.  We can then write
Eq.(\ref{dpdec2}) as
\begin{equation}p^C_{b_{up}}(b_c)
\approx {2\pi \over 9}\left ( {Z_{T}e\over \hbar
vb_c}\right)^2\int d\varepsilon_k
2\left (2 \varpi K_{1}\left(  \varpi\right)\right)^2
 {dB(E1,\varepsilon_k)\over d\varepsilon_k}.
\label{dpdec3}\end{equation}
 since for the case studied in this paper the explicit form of B(E1) is
\begin{equation}{dB(E1,\varepsilon_k)\over d\varepsilon_k}= C^2S {\frac{m_n}{\hbar^{2}k}}
 (C_i{\beta}_{1}Z_{P}e)^2{6\over \pi^2}{k^4\over (k^2+\gamma^2)^4},\label{dBde}\end{equation}
which  is the consistent with Eq.(6.5) of Ref.\cite{he} or Eq. (7.24) of \cite{carlos} and
it is the explicit form of the B(E1) obtained for an s-initial state using the asymptotic form of the
wave function.

Finally using Eq.(\ref{I}) in Eq.(\ref{alpha}) and comparing with Eq.(\ref{dpdec3}) we obtain that
$\alpha =-{p^C_{b_{up}}/ 2}$ in the high energy,  straight line trajectory  limit. However we remark that  our probability Eq.(\ref{dpdec}) is more accurate than
 the high energy limit of  $\alpha$ of Eq.(\ref{alpha}) since it contains also the term representing
longitudinal excitations, proportional to $K_0$ and  to $k_z$, the parallel component of neutron
momentum.

\end{document}